\documentclass[twocolumn,aps,prl,superscriptaddress]{revtex4-2}

\usepackage{amsmath}
\usepackage{graphicx}
\usepackage{MnSymbol}
\usepackage{color}
\usepackage{comment}

\begin{document}

\title{Crack and pull-off dynamics of adhesive, viscoelastic solids}
\author{Martin H. Müser}
\affiliation{Dept. of Materials Science and Engineering, Saarland University, 66123 Saarbrücken, Germany}
\author{Bo N. J. Persson}
\affiliation{PGI-1, FZ Jülich, Germany}
\affiliation{MultiscaleConsulting, Wolfshovener str 2, 52428 Jülich, Germany}

\begin{abstract}
When quickly detaching an elastomer from a counterface, viscoelasticity dramatically increases the perceived adhesion relative to its adiabatic or equilibrium value.
Here, we report simulations on the sticking contact between a rigid cylinder and a viscoelastic half space
revealing a maximum in the work of adhesion at intermediate pull-off velocities. 
Maximum tensile forces yet increase monotonically with the pull-off speed and the crack-tip speed in accordance with the Persson-Brener approach.
As predicted theoretically, the fracture mode transitions 
from interfacial crack propagation to quasi-uniform bond breaking with increasing range of adhesion. 
\end{abstract}

\maketitle

We all know since childhood that the pain experienced when tearing off a bandage is small when pulling either very slowly or very  quickly. 
In between these two limits, it hurts.
Surely, one important reason for this phenomenon is that  breaking an adhesive, viscoelastic interface is crucially affected by the interplay of the interfacial energy, the maximum tension of the media in contact, the frequency dependence of their mechanical properties, and the pull-off velocity~\cite{Knauss2015IJF,Creton2016RPP}.
%
Similar comments can be made about the rupture and wear of rubber~\cite{Gent1996L,Persson2005JPCM} as well as the adhesion, cohesion, and friction involving related elastomers including, for example, pressure-sensitive adhesives~\cite{Villey2015SM}, tapes~\cite{Afferrante2016JMPS}, or cartilage~\cite{Han2020JMBBM}.
%
Unfortunately, even the most elementary linearly viscoelastic, adhesive interfaces (for which fibrillation, cavitation, and other complex phenomena that  matter for the bandage example~\cite{Persson2005JPCM,Villey2015SM} can be neglected) defy a simple description of their dynamics. 

\begin{figure*}[btp]
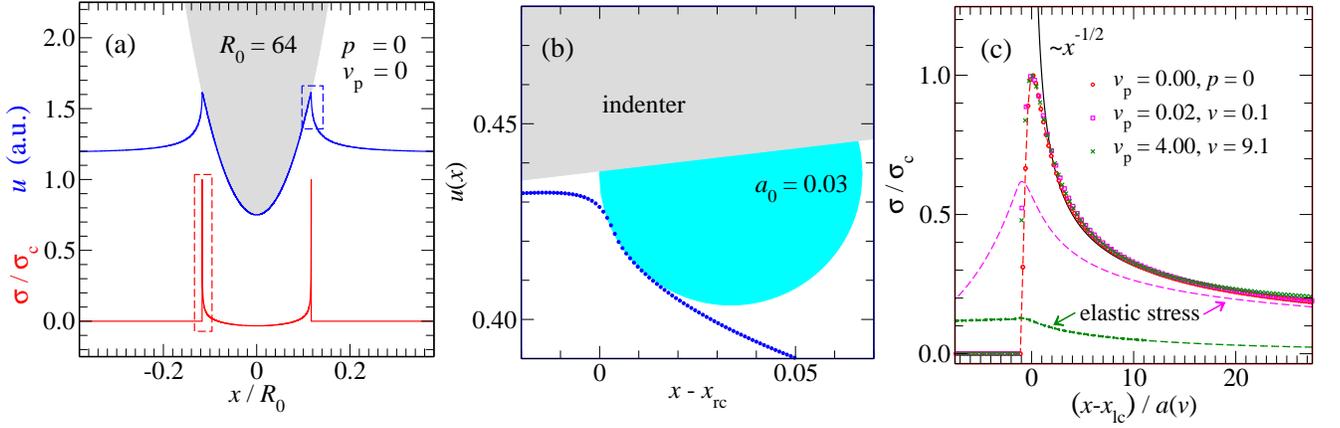

\centering
\begin{minipage}{0.32\textwidth}
\includegraphics[width=1\textwidth,angle=0]{equilKonf.eps}
\end{minipage}
\begin{minipage}{0.32\textwidth}
\includegraphics[width=1\textwidth,angle=0]{circle2.eps}
\end{minipage}
\begin{minipage}{0.32\textwidth}
\vspace*{2mm}
\includegraphics[width=1\textwidth,angle=0]{stressSym.eps}
\end{minipage}
\caption{
(a) Displacement field $u(x)$ (blue) and interfacial stress $\sigma$ (red) in units of the maximum stress $\sigma_\textrm{c}$ for the system with $R_0 = 64$ and $E_1/E_0= 100$.
Zooms into (b) contact geometry relative to the right crack tip located at $x_\textrm{rc}$ and (c) interfacial stress relative to the left crack at $x_\textrm{lc}$.
The elastic stress fields, defined as the equilibrium elastic stress for a fixed $u(x)$, are shown for comparison in panel (c).
The lateral coordinate $x$ is normalized differently in different panels. 
}
\label{fig:realSpace}
\end{figure*}

The critical quantity in a viscoelastic fracture problem is the energy per unit area, $G(v)$, needed to advance a crack by a unit area as a function of the crack tip speed $v$. 
Traditionally~\cite{Knauss2015IJF,Schapery1975IJFa,Schapery1975IJFb,Barber1989PRA,Hui1992JAP,Haiat2003JMPS,Greenwood2004JPD,Greenwood2007JPD}, the attempt is made to determine $G(v)$ 
from the solution of a self-consistent equation, which first needs to be derived for each combination of a given frequency-dependent elastic modulus $E(\omega)$ and  cohesive-zone model (CZM).
The latter states how adhesive or cohesive stress changes locally with the interfacial separation, or, gap $g$.
However, as pointed out by de Gennes~\cite{deGennes1996L}, certain universal features should apply given that the stress near crack tips generally obeys $\sigma(r) = K/\sqrt{2\pi r}$ a small distance $r$ away from the crack tip~\cite{Freund1998book}, where $K$ is called the stress-intensity factor, see also Fig.~\ref{fig:realSpace}(c). 
In the immediate vicinity of a fast moving crack and very far away from it, the contact mechanics are similar to that of an adiabatically moving crack, however, assuming the high- and low-frequency elastic modulus, $E_1$ and $E_0$, at small and large $r$, respectively. 
Unfortunately, the interesting, non-trivial intermittent region is where most energy can be dissipated, whereby this region may predominantly account for the increase of $G(v)$ compared to its quasi-static or adiabatic value $G_0$.

To quantify the viscous energy loss for a steadily moving crack, Persson and Brener~\cite{Persson2005PRE} argued that the stress singularity near the crack tip is cut-off by the local maximum tension $\sigma_\textrm{c}$. 
This made them introduce 
a speed-dependent wave number cutoff $q_\textrm{c}(v)$, above which the elastomer no longer noticeably deforms. 
The cut-off reveals itself experimentally through a blunting of the crack tip at large crack-tip speeds. 
It can be obtained through the self-consistent equation
\begin{equation}
\label{eq:crackTipRadius}
   q_\textrm{c}(v) = q_0\, \left\{ 1 - 
   I\left(v\,q_\textrm{c}(v)\right) \right\},
\end{equation}
where the static cut-off wave number $q_0 = q_\textrm{c}(0)$, whose relation to other characteristic distances and wave numbers is discussed in the supplemental material (SM), is the only adjustable parameter, and where 
\begin{equation}
    I(\omega) = 
    \frac{2}{\pi} \int_0^1\!\textrm{d}x\, \frac{\sqrt{1-x^2}}{x}\mathrm{Im}\left\{\frac{E_0}{E(x \omega)}\right\}.
\end{equation}
From this, $G(v)$, which turns out inversely proportional to $q_\textrm{c}(v)$, can be deduced through
\begin{equation}
\label{eq:GofVandA}
    G(v)\, q_\textrm{c}(v) = G_0\, q_\textrm{c}(0).
\end{equation}
An important advantage of the Persson-Brener approach is that a relatively simple self-consistent integral equation needs to be solved, in which $E(\omega)$ can be arbitrarily complex without obstructing the calculation of $G(v)$. 
Despite this benefit, the Persson-Brener approach~\cite{Rodriguez2020APS} yields a similar speed-dependence of the fracture energy as that determined from traditional solutions~\cite{Hui1992JAP,Greenwood2004JPD,Afferrante2021arX} of simple rheological models.

Neither the traditional nor the Persson-Brener approach have hitherto been verified against rigorous numerical solutions over a meaningfully large parameter range.
%
One purpose of this article is to fill this gap.
A further and equally important issue addressed here is the question how viscoelastic crack propagation affects the snap-off dynamics in single-asperity contacts. 
This includes a test of the prediction~\cite{Persson2003W,Gao2004PNAS} that the fracture mode changes from interfacial crack propagation to quasi-uniform bond breaking at small scales and an analysis of how the work of separation depends on the pull-off velocity. 

A convenient set-up for our analysis is the contact between a rigid cylinder, which we approximate with a parabola having a (varying) radius of curvature $R_0$, and a viscoelastic half space.
During pull-off, the boundary lines between the contact and non-contact region constitute two linear cracks, which propagate inwards until an elastic snap-off instability occurs.
For this situation analytical results relate the pull-off force to the crack propagation energy $G(v)$, which depends on the crack tip speed $v$ at the point of snap-off~\cite{Barquins1988JA,Chaudhury1996JAP,Liu2015SM}.

The model studied in this work, see also Fig.~S1 in SM, assumes a regular three-element viscoelastic model, for which
\begin{equation}
    \frac{E_0}{E(\omega)} = \frac{E_0}{E_1}+\left (1-\frac{E_0}{E_1}\right ) \frac{1}{1-i\omega \tau},
\end{equation}
and a recently proposed CZM $V(g)$~\cite{Wang2021L}, 
where $V(g)$ is zero if $g$ exceeds the cut-off gap $g_\textrm{c}$ and
\begin{equation}
    V(g) = -\Delta\gamma\, \times
    \begin{cases}
\left [1-(\pi\,g/g_\textrm{c})^2/2\right ] \ & \textrm{if } g<0 \\
\left [ 1+\cos(\pi\,g/g_\textrm{c})\right ]/2 & \textrm{if } 0 \le g < g_\textrm{c} 
\end{cases}
\end{equation}
otherwise.
Here, $\Delta\gamma$ is the interfacial binding energy gained when  cylinder and elastomer touch ($g=0$). 
An advantage of the employed CZM over  commonly used Dugdale-type models are that ours is twice differentiable, as real interactions are, whereby numerical solutions of the dynamics are quite robust.

Simulations were conducted using a house-written Green's function molecular dynamics (GFMD) code, which has been described numerous times before, see, e.g., Appendix~2 in Ref.~\cite{Prodanov2014TL}.
However, the used propagator was changed in order to reflect the dynamics of the standard three-element model leading to a similar approach as that pursued by Bugnicourt \textit{et al.}~\cite{Bugnicourt2017TI}.
To improve numerical stability, the interfacial stress and its time derivative, that is, the r.h.s. of Eq.~(4) in Ref.~\cite{Bugnicourt2017TI}, were low-pass filtered as described elsewhere~\cite{Sukhomlinov2022ASSA}.

The length of the periodically repeated simulation cell was generally set to $L = 4\,R_0$, where $R_0$ took the values $R_0 = 1$, 8, and 64. 
Note that three variables can be used to define the unit system. 
Throughout this work, we assume a unit system in which the contact modulus $E_0^*\equiv E_0/(1-\nu^2)$, $\tau$, and the smallest $R_0$ define the units of stress, time, and length, respectively. 
Here, $\nu$ is the Poisson ratio, which is assumed to not depend on frequency. 
Real units can be produced by setting, e.g., $E_0^* = 5$~MPa, $\gamma = 50$~mJ/m$^2$, in which case the unit of length would be 10~nm.
From a continuum prospective, it might be more meaningful to state the Tabor parameter, which would read $\mu_\textrm{T} = \sqrt[3]{R_0\sigma_\textrm{c}^3/({E_0^*}^2\Delta \gamma)}$,
if the ratio $\Delta \gamma/\sigma_\textrm{c}$ was used as the range of interaction in the common definition of $\mu_\textrm{T}$.
With $\sigma_\textrm{c} = \pi\,\Delta\gamma/(2\,g_\textrm{c})$, the Tabor parameters realized in this study would range from $\mu_\textrm{T}\approx 4$ for $R_0 = 1$ to $\mu_\textrm{T}\approx 16$ for $R_0 = 64$, which could be classified as medium to short-ranged adhesion. 
In comparison, Afferante and Violano studied effective surface energies in viscoelastic Hertzian contacts in the limit of long-range interaction, i.e., for $\mu_\textrm{T} \approx 1/3.85$, and the fixed ratio $E_1/E_0 = 10$ in a compelling, recent study~\cite{Afferrante2021arX}.

Space was discretized into elements with lateral dimension $\Delta x = 1/1024$, which was fine enough to prevent finite-discretization induced instabilities, also known as lattice trapping, 
for the current choice of parameters.
They were $E_0^* = 1$, $\Delta \gamma = 0.01$, and $g_\textrm{c} = 0.0175156$, so that $\max(V''(g)) = -\min(V''(g)) = (\pi/g_\textrm{c})^2\Delta\gamma/2 $ was exactly one tenth of the maximum static elastic stiffness $\kappa_\textrm{max} = q_\textrm{max} E^*_0 / 2 $ with $q_\textrm{max} = \pi/\Delta x$.
This means that quasi-static continuum mechanics of short-range adhesion is closely approached on lateral lengths exceeding $\mathcal{O}(10~\Delta x)$ but inadequate at smaller scales.

To further illuminate the model, Fig.~\ref{fig:realSpace}(a) shows the overall contact geometry and the stress field for a force-free static contact and panel (b) a zoom into the displacement field showing our determination of the static crack-tip radius $a_0$.
Moreover, Fig.~\ref{fig:realSpace}(c), see also Fig.~S2 in SM, confirms that the interfacial stress $\sigma_\textrm{int}$ in the vicinity of the crack tip obtained at different $v$ can be superimposed when scaling the distance from the crack tip with the ratio $q(v)/q_0$ deduced from Eq.~(\ref{eq:crackTipRadius}). 
For $v = 0$, the elastic stress, $\sigma_\textrm{el}$ defined as the inverse Fourier transform of $qE^*_0\tilde{u}(q)/2$, coincides with $\sigma_\textrm{int}$.
At intermediate $v$, $\sigma_\textrm{el}$  is still relatively close to $\sigma_\textrm{int}$ in the immediate vicinity of the crack tip and approaches it asymptotically at large distances from the crack tip.
However, elastic and interfacial stress differ substantially at large $v$. 
Since relaxation is driven by the difference between elastic and interfacial stress, dissipation occurs predominantly far away from the cracks in the latter case. 

\begin{figure}[hbt]
    \centering
    \includegraphics[width=0.47\textwidth]{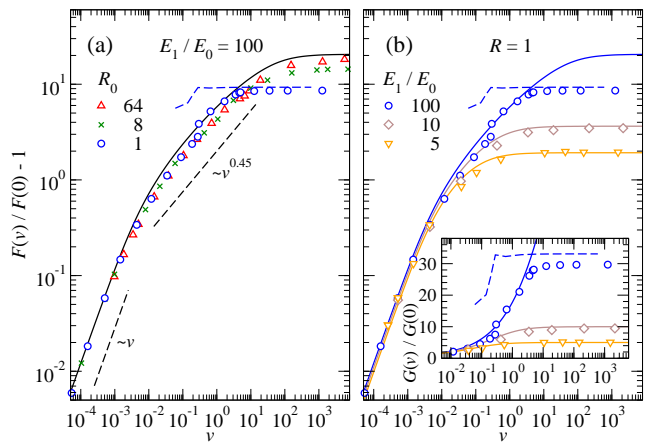}
    \caption{Relative pull-off force increase, $F_\textrm{p}(v)/F_\textrm{p}(0)-1$, as a function of crack-speed velocity $v$ for (a) fixed $E_1/E_0=100$ and varying radius of curvature $R_0$ and  (b)  fixed $R_0=1$ and varying $E_1/E_0$. Symbols and lines reflect simulation and theoretical results, respectively. Blue dashed lines reflect upper bounds to $F_\textrm{p}$. The inset in panel (b) shows the ratio $G(v)/G_0$. }
    \label{fig:pullForce}
\end{figure}

Analytical results predict that the maximum tensile force, also called the pull-off force, $F_\textrm{p}$, satisfies $F_\textrm{p} = (27\,\pi\, G^2\,E^*_0\,R_0/16)^{1/3}$~\cite{Barquins1988JA,Chaudhury1996JAP,Liu2015SM}.
Treating the breaking of the adhesive bonds between the solids as the propagation of an opening interfacial crack also at the moment of pull-off, we therefore expect
\begin{equation}
    \frac{F_\textrm{p}(v)}{F_\textrm{p}(0)} = \left(\frac{G(v)}{G_0}\right)^{2/3}, 
\end{equation}
where $v$ is the crack-tip velocity at the moment when the normal force reaches its maximum. 
In fact, Fig.~\ref{fig:pullForce} reveals close agreement between simulation and theory for how the pull-off force increases with pulling speed.
The static pull-off forces, $F_\textrm{p}(0)$, needed to accurately normalize $F_\textrm{p}(v)$ were deduced from mass-weighted GFMD simulations~\cite{Zhou2019PRB} using very small $v_\textrm{p}$.
They deviated at most by 0.1\% from the just-stated, quasi-static continuum expression for $F_\textrm{p}$.
%

Persson-Brener theory (full lines in Fig.~\ref{fig:pullForce}) match within the numerical precision in the linear-response regime at small velocities.
This linear-response regime arises as a consequence of how the modeler (or nature!) discretizes the elastic manifold.
For coarse discretization, lattice pinning occurs so that instabilities become unavoidable~\cite{Holland1998PRL}, which in turn lead to Coulomb friction.
However, at fine discretization, ``atoms'' move continuously at all times under adiabatic driving and Stokes damping arises automatically.
Power-law scaling of the damping force would only be expected down to infinitesimal small velocities right at the critical point (in the absence of thermal noise) separating the Stokes from the Coulomb regime~\cite{Muser2002PRL}. 
There will still be extended velocity regimes, in which sub-linear small-velocity corrections to either $G(v)$ or $F(v)$ arise even if $G(v)-G(0)$ and thus $F(v)-F(0)$ ultimately cross over to Stokes, whenever $1/\sqrt{\vert \Delta r \vert}$ stress singularity near crack tips extends down to small but not atomic scales.

Differences between Persson-Brener theory and simulations reach 30\% at intermediate velocities and decrease again for large tip radii at large $v$, where the viscoelastic fracture energy factor $G(v)/G_0$ plateaus close to the predicted value of $E_1/E_0$. 
The latter ratio can be directly deduced from the theory by combining Eqs.~(\ref{eq:crackTipRadius}) to (\ref{eq:GofVandA}) and by realizing that $I(\omega\to\infty)\to 1$.
The close match between theory and simulation is an interesting result in its own right, also because the theory assumes steady-state crack propagation, while in reality, the crack-tip speed is not constant at fixed pull-off velocity.
Moreover, even better agreement must be expected for systems with a broad distribution of relaxation times, as the sharp-wavenumber-cutoff approximation in the Persson-Brener approach should be most inaccurate for the three-element model with a single relaxation time.

%

Theory and simulation differ significantly in Fig.~\ref{fig:pullForce} for small tip radii when $v$ and $E_1/E_0$ are both large.
This is due to the transition of the failure mode from crack propagation to quasi-uniform bond breaking, which was proposed to occur at small scales~\cite{Persson2003W,Gao2004PNAS}. 
The argument for the phenomenon is that the tensile load in a finite contact should be roughly limited by the product of the maximum tensile stress and the contact width $w$.
In fact, the dashed lines reflecting this estimate match the large-velocity limit for the $R_0 = 1$, $E_1/E_0 = 100$ system quite well if the value for $w$ is the one observed in the simulations at the moment of maximum tensile force. 

\begin{figure}[hbt]
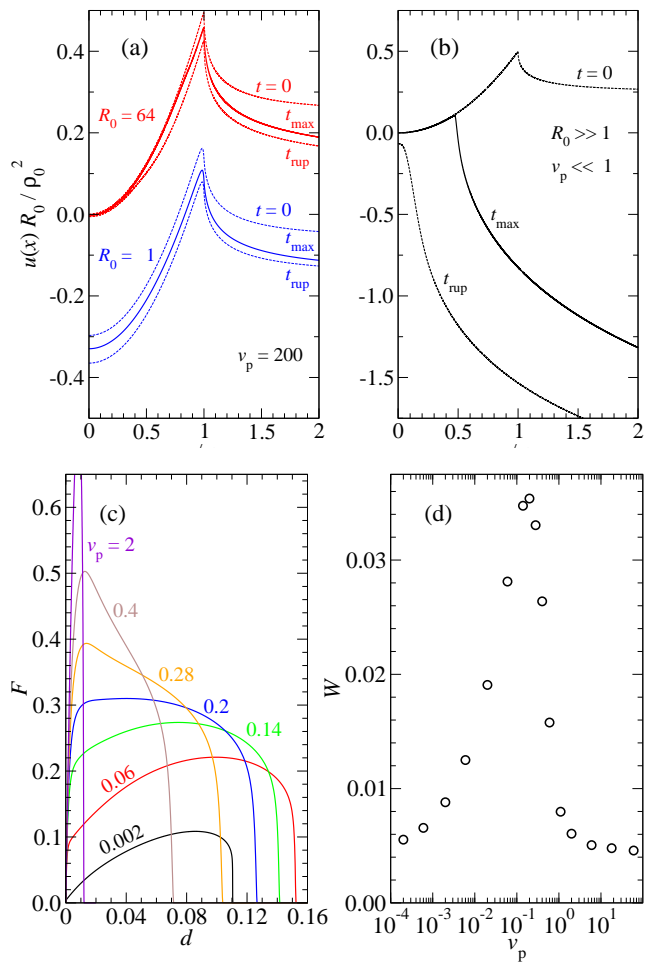

    \centering
    
\includegraphics[width=0.47\textwidth]{sizeEffectB.eps}
    
    \includegraphics[width=0.47\textwidth]{pullDyn.eps}
    \caption{(a) Displacement fields in static equilibrium at zero applied force ($t = 0$), when the tensile force is maximum ($t = t_\textrm{max}$), and at the moment of final rupture $t = t_\textrm{rup}$ in the limit of large velocities for contact radii $R_0 = 64$ (red curves) and $R_0 = 1$ (blue curves).
    (b) Similar as (a) but for small pulling velocities and large $R_0$. 
    (c) Force-displacement curves at different velocities for the $R_0=1$ tip and (d) its velocity-dependent work of separation $W$.}
    \label{fig:pullDyn}
\end{figure}

The suggested quasi-uniform bond breaking is also borne out from the displacement fields shown in Fig.~\ref{fig:pullDyn}(a): 
at large $v_\textrm{p}$,
the $R_0 = 64$ contact evidently fails by crack propagation  while the displacement field moves almost homogeneously during failure for $R_0 = 1$.
Specifically, for $R_0 = 1$ contact is already lost at $r = 0$ when the force reaches its maximum, while for $R_0 =64$, there is still contact near the origin directly after the moment of final rupture, which we define as the point in time right at which the tensile-load displacement curve assumes its most negative slope. 
At small velocities, all contacts studied here break in a similar way as shown for large $R_0$ in Fig.~\ref{fig:pullDyn}(b). 
This is because the Tabor parameter is greater than unity even for $R_0 = 1$ so that the adiabatic tip retraction is close to the continuum limit~\cite{Maugis1992JCIS,Muser2014B}.

The blue, dashed lines in Fig.~\ref{fig:pullForce} also reveal that the slope of the critical width $w(v)$, defined as the width of the contact at the moment of maximum force, changes discontinuously at a certain crack-tip velocity, which, however, is unrelated to the transition in the failure mode.
Similar discontinuities in $w'(v)$  occur for all investigated systems.
As no theory apparently predicts this transition,  all currently existing analytical approaches to the crack-tip problem could be argued to be approximate. 

To further illuminate the pull-off dynamics,  Fig.~\ref{fig:pullDyn}(c) shows various load-displacement curves $F(d)$ for $R_0 = 1$.
Their shape changes indeed abruptly near $v_\textrm{p} = 0.2$, for which $F(d)$ has a very flat maximum. 
At that pull-off velocity, $d_\textrm{max}$---the vertical distance moved to reach the maximum force---changes quite quickly from a value of order $d_\textrm{max}(v\to 0) \approx 0.08$ to $d_\textrm{max}(v\to \infty) \approx g_\textrm{c}/2$, where the CZM assumes its maximum tensile stress. 

Owing to the small forces needed to separate surfaces adiabatically and the small $d_\textrm{max}$ needed to break the contact at large $v_\textrm{p}$, the work of separation $W$ turns out small in both limits.
In between, viscous dissipation is largest leading to a pronounced maximum in $W$, which is shown in Fig.~\ref{fig:pullDyn}(d). 
It can be said to arise, because a tensile force close to $F_\textrm{p}(v\to \infty)/2$ acts over relatively large pulling distances. 
Similar trends are found for larger system size when keeping $R_0$ and all other parameters constant as well as for regular Hertzian tips, for which $W$ does not have logarithmic system-size corrections, see also Fig.~S4 in  SM.
Also flat, circular punches show a maximum in $W(v_\textrm{p})$ at intermediate $v$, since both high- and low-velocity separation are easily found to be $2\,\gamma$ times punch area, at least as long as the range of adhesion remains short-ranged even when the elastomer assumes its high-frequency modulus.

The maximum in $W$ might appear counterintuitive, since $G(v)$ monotonically increases with $v$.
%
However, only a small fraction of the initial contact is broken when the normal force assumes its maximum at large pulling velocities.
Past that point, the crack tip velocity can quickly increase with time, in particular right before snap-off, as revealed in Fig.~S3 in SM, 
so that the viscoelastic crack-propagation theory, which assumes slowly changing crack-tip speeds, no longer holds. 
A maximum loosely related to ours in $W(v_\textrm{p})$ occurs in the work needed to roll a cylinder a unit distance over a viscoelastic half space as a function of rolling speed~\cite{Hunter1961JAM,vanDokkum2019MSMSE}.

Our calculation of the work of separation does not include the energy loss due to the (visco-) elastic coupling between the center-of-mass mode of the elastomer's surface facing the indenter and its other surface, which is typically driven in laboratory experiments.  
However, only the work of separation due to finite $q$ is dissipated in the vicinity of the contact so that we expect pull-off induced near-surface heating to be largest at intermediate velocities. 
This effect should also hold for interfaces that are more complex than the one investigated here.
For example, the product of maximum stress and the time during which nerves in the vicinity of hair roots are exposed to large forces should be maximal at intermediate pull-off velocities, which might explain the  sudden, but quickly decaying pain that we experience when pulling off a bandage with an intermediate velocity.
%


MHM acknowledges useful discussions with Sergey Sukhomlinov and Christian Müller.

\bibliographystyle{plain}
\bibliographystyle{unsrt}

\end{document}


\title{Supplementary materials: Crack and pull-off dynamics of adhesive, viscoelastic solids}
\author{Martin H. Müser}
\affiliation{Dept. of Materials Science and Engineering, Saarland University, 66123 Saarbrücken, Germany}
\author{Bo N. J. Persson}
\affiliation{PGI-1, FZ Jülich, Germany}
\affiliation{MultiscaleConsulting, Wolfshovener str 2, 52428 Jülich, Germany}

\maketitle

\section{Details on model and set-up}

\begin{figure}[hbp]
\centering
\includegraphics[width=0.633\textwidth,angle=0]{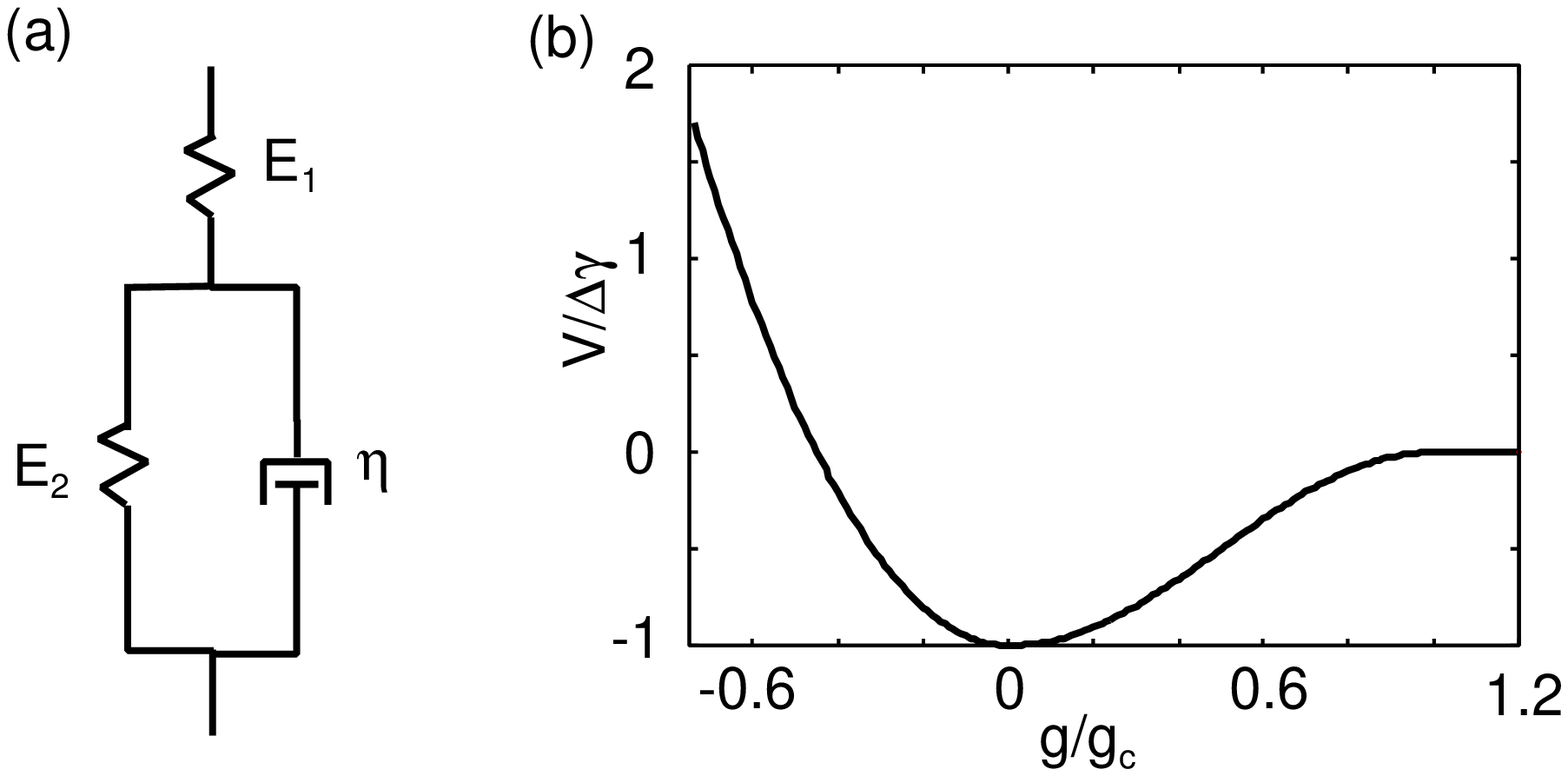}
\caption{
(a) Three-element viscoelastic model.
The low-frequency modulus $E_0 \equiv E(\omega=0) =E_1 E_2/(E_1+E_2)$,
the high-frequency modulus $E(\infty)=E_1$,
and the viscosity $\eta=1/\tau$ are indicated.
(b) The wall-wall interaction potential (per unit surface area) used in the numerical simulations. 
}
\label{rheologypic2.eps}
\end{figure}

Fig.~\ref{rheologypic2.eps}(a) shows the used three element model.
$E_1$ can be easily seen to represent the high-frequency modulus $E(\omega\to\infty)$, while $E_0 = E_1\,E_2/(E_1+E_2)$ represents the small-frequency modulus $E(0)$.
Fig.~\ref{rheologypic2.eps}(b) depicts the cohesive zone model used in the numerical simulation. 
The curvature of $V(g)$ is most negative at the cut-off distance $g_\textrm{c}$, however, $V(g)$ and $V'(g)$ are continuous everywhere. 

\section{Characteristic wave numbers and distances}

Quite a few different characteristic distances and wave numbers occur in the contact problem treated in the main paper. 
They should be generally related by conversion factors of order unity times $1/(2\pi)$ to $2\pi$. 
It may be beneficial to introduce them and to identify their close-to-exact relation for the used cohesive-zone model (CZM). 
We expect their values to be similar for other CZMs so that they remain close to unity and no longer extend all the way to $2\pi$.

The first wave vector to be introduced for our CZM is the one at which the elastic stiffness $\kappa_\textrm{el} = qE^*_0/2$ is equal to the maximum negative curvature of the interaction potential, i.e.,
\label{eq:qs}
\begin{eqnarray} \label{eq:qsDef}
    q_\textrm{s} & = & 2\,\max(-V''(g))/E^*_0.
\end{eqnarray}
For our CZM, 
$\sigma_\textrm{c} = (\pi/2)\Delta\gamma/\,g_\textrm{c}$ and
$\max(-V''(g))= (\pi/2)^2\Delta\gamma/g_\textrm{c}^2$ so that
\begin{equation}
    q_\textrm{s} = \frac{2\,\sigma_\textrm{c}^2}{E^*_0\,\Delta\gamma}.
    \label{eq:qsGeneral}
\end{equation}
Modes with wave vectors $q \ll q_\textrm{s}$ will always behave as in the continuum / short-range-adhesion limit, while those with $q \gg q_\textrm{s}$ are too stiff to be distorted by the interfacial  interactions. 
Thus, the wave-number cutoff $q_\textrm{c}$ used in the Persson-Brener theory must be of order $q_\textrm{s}$.
A motivated guess for the used conversion factor $\alpha_\textrm{cs} \equiv q_\textrm{c} / q_\textrm{s}$ is provided further below. 

We note in passing that Eq.~(\ref{eq:qsDef}) is not directly applicable to CZMs in which $V(g)$ is not twice differentiable or assumes a ``funky'' shape. 
However, for conventional CZMs, like the Dugdale model, Eq.~(\ref{eq:qsGeneral}) should remain applicable within deviations of order~10\%. 

We are also interested at a characteristic stiffness distance $x_\textrm{s}$~\footnote{The term ``stiffness distance'' is somewhat odd but due to the fact that the words characteristic, cohesive, contact,  crack, critical, and cross-over all start with the letter c and three out of those six words even start with cr, the need for unconventional name giving arose.}  from the crack tip in real space at which the the stress closely approaches the asymptotic $\sigma(x) = K/\sqrt{2\pi \vert x \vert}$ behavior.
As argued in the main text, this distance is of order $x_\textrm{s} \equiv 2\pi/q_\textrm{s}$.
A more precise estimate can be obtained by realizing that the real stress field inside the contact should roughly follow
\begin{equation}
\label{eq:sigmaInContact}
    \sigma(x>0) \approx \frac{\sigma_\textrm{c}\sqrt{x_{K}}}{\left({x^2+x_{K}^2}\right)^{1/4}},.
\end{equation}
where $x_\textrm{K}$ is again of order $x_\textrm{s}$.
Other CZMs will lead to other $\sigma(r)$ dependencies than ours.
However they will all obey $\sigma(0) = \sigma_\textrm{c}$, $\sigma'(0)=0$, and $\sigma(x\to\infty) = K/\sqrt{2\pi\,x}$, which Eq.~(\ref{eq:sigmaInContact}) reflects. %
From Eq.~\eqref{eq:sigmaInContact}, it follows that $K = \sigma_\textrm{c}\sqrt{2\pi x_K}$.
Comparing this result to Eq.~(19) in Ref.~\cite{Persson2005PRE}, it follows that $x_K = a(0)$.
However, we note that the value that we used for $a_0$ was adjusted so that the Stokesian friction for the $R_0 = 64$ tip was accurately reproduced.
This was achieved with the numerical value of $a_0 = 0.019$.
This value is close to but slightly less than the static crack tip radius, which is roughly $0.03$.

\begin{figure}[hbt]
    \centering
    \includegraphics[width=0.5\textwidth]{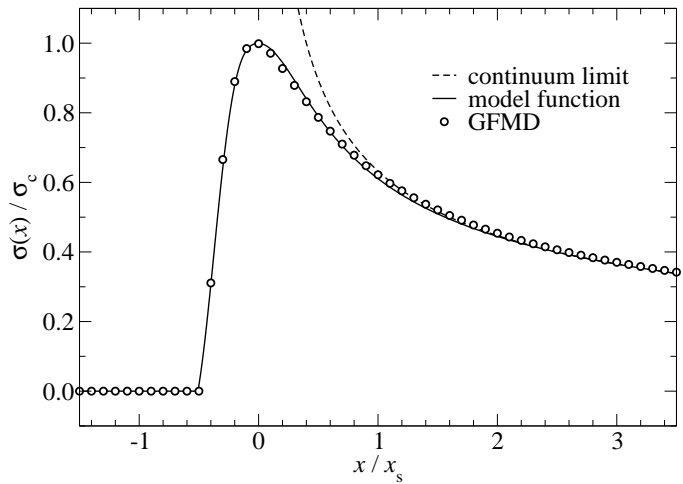}
    \caption{\label{fig:stressNormA} Equilibrium stress profile for the $R_0=60$ tip at zero external load.
    Each GFMD grid point is represented by a circle. 
    The dashed line shows the continuum limit.
    The full line reflects Eq.~\eqref{eq:sigmaInContact} for positive $x$ and is meant to guide the eye for negative $x$.
    }
\end{figure}

The parameter $x_\textrm{K}$ can be gauged from the computed stress field as demonstrated in Fig.~\ref{fig:stressNormA}(a). 
The model function used to determine $K$ and $x_K$, i.e., the full line drawn for positive $x$ in Fig.~\ref{fig:stressNormA}(a), was obtained using $x_\textrm{K} = 0.4~x_\textrm{s}$.
It includes the stress-field from the other crack opening and the Hertzian stress field counterbalancing the two stress fields from the two crack tips to yield zero total pressure. 
\vspace*{1cm}

\section{Analytical solution for the three-element model}

For the employed three-element method, the integral $I(\omega)$ defined in the main paper becomes~\cite{Persson2005PRE}
\begin{equation}
    I(\omega) = \left(1-\frac{E_0}{E_1}\right)\,\frac{2}{\pi}\,\int_0^1\!\mathrm{d}x \, \sqrt{1-x^2}\,\frac{\omega\tau}{1+(\omega\tau x)^2}.
\end{equation}
It has the solution
\begin{equation}
I(\omega) = \left(1-\frac{E_0}{E_1}\right)\frac{\sqrt{1+(\omega\tau)^2}-1}{\omega\tau}.
\end{equation}
As a consequence, the self-consistent equation to be solved for a standard, three-element half space becomes
\begin{equation} \label{eq:selfConstApp}
    \frac{q(v)}{q_0} = 1 -
    \left(1-\frac{E_0}{E_1}\right)
    \frac{\sqrt{1+q^2(v)v^2\tau^2}-1}{q(v)\,v\,\tau}
\end{equation}
after substituting $\omega = v\,q_\textrm{c}(v)$.
In principle, this is a quadratic equation in $q(v)$ and therefore analytically solvable.
However, the coefficients are cumbersome so that we found a self-consistent solution of Eq.~\eqref{eq:selfConstApp} for $q(v)$ to remain most convenient. 

\section{Analytical solution for cylinder at depinning}

The analytical solution for a quasi-static cylinder in contact with a short-range adhesive, linearly elastic solid reads~\cite{Barquins1988JA}
\begin{equation}
    F= \pi\,E^*_0 b^2 /(4 R) - (2 \pi E^*_0 b G(0))^{1/2}.
\end{equation}
For a dynamic system, we replace $G(0)$ with $G(v)$.
The tensile force $F$ then becomes maximal when
\begin{equation}
b = \left( \frac{2\,G(v)\,R^2}{\pi\,E^*} \right)^{1/3}.
\label{eq:halfWidth}
\end{equation}
leading to the pull-off force of
\begin{equation}
F_\textrm{p} = \left( 27\,\pi\,E^*R/16\right)^{1/3},   
\end{equation}
which was already mentioned in the main text.

\section{Work of separation}
\begin{figure}[hbtp]
    \centering
    \includegraphics[width=0.66\textwidth]{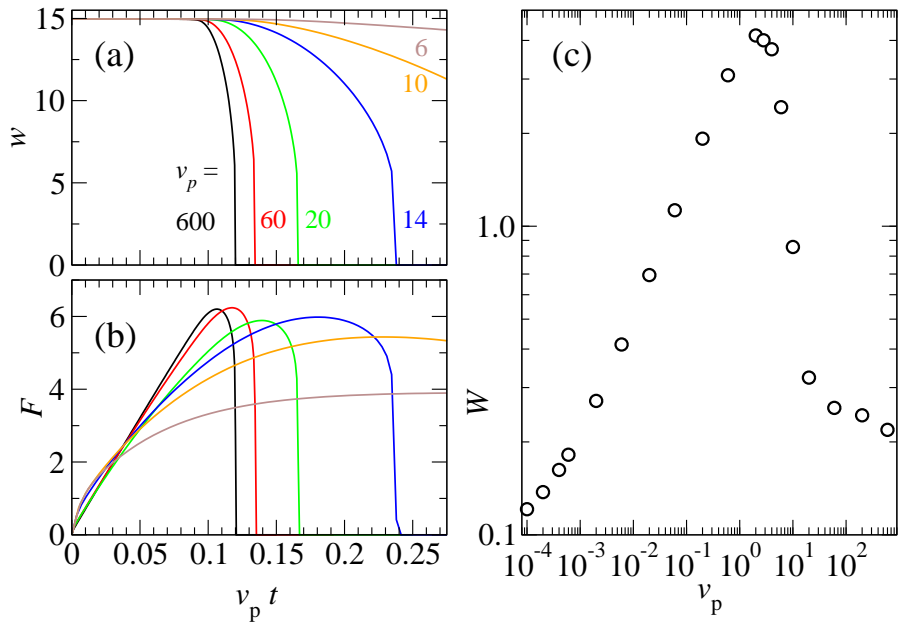}
    \caption{(a)~Contact width $w$ width and (b)~tensile force $F$ as a function of the moved distance $v_\textrm{p}\,t$ for the $R_0 = 64$ system.
    (c)~Work of separation as a function of pulling velocity $v_\textrm{p}$.}
    \label{fig:crackR64}
\end{figure}

In the main manuscript, we argue the crack-propagation theory to break down at large pulling velocities after the tensile force assumed its maximum so that it cannot be used directly to estimate the work of separation. 
To corroborate this claim, we show how tensile force and the crack-tip position evolve with the slid distance for the $R_0 = 64$ system in Fig.~\ref{fig:crackR64}.
It can be seen that dynamics become quasi-discontinuous at the moment when the tensile forces assume their maximum, at which point the contact radius is barely decreased relative to its initial value. \vspace*{1cm}

\section{Geometries beyond cylinders}
In the main manuscript, we report having observed local maxima in $W(v_\textrm{p})$ for adhesive elastomers also for geometries other than cylinders. 
To support this claim, we briefly discuss the regular flat punch and also present numerical results for a regular Hertzian indenter. 

For the flat punch interacting through small-range adhesion, the contact area remains unchanged until the tensile force reaches its maximum, at which point a crack starts propagating.
For a flat punch with radius $\rho_0$, the force-distance relation reads $d = F/(2\,\rho_0\,E^*)$, while the pull-off force is given by $F_\textrm{p} = \sqrt{8\,\pi\,E^*\,\Delta\gamma\,\rho_0^3}$.
Thus, in the adiabatic case, the work of separation, $W = \int_0^{F_\textrm{p}}\!\mathrm{d}{F} \, d(F) = 2\Delta\gamma$, does not depend on $E^*$.
This is why $W$ turns out identical for the high- and low-frequency modulus, i.e., in the limits $v \to 0$ and $v\to\infty$.
Since $W(v)$ evaluated at slightly positive $v$ automatically exceeds $W(0)$, there must be a maximum in the work of adhesion between the limits of infinitesimally small and infinitely large velocity. 
The correctness of these conclusions was validated numerically. 
A detailed analysis of the depinning of a flat punch from a viscoelastic foundation is currently under preparation. 

\begin{figure}[hbtp]
    \centering
    \includegraphics[width=0.5\textwidth]{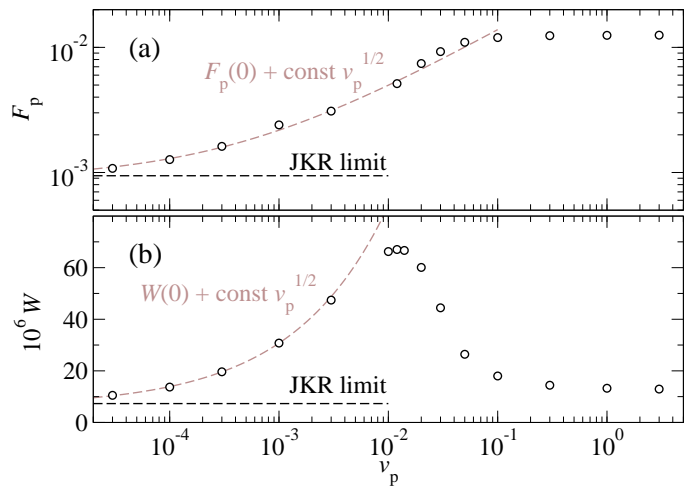}
    \caption{(a) Depinning force $F_\textrm{p}$ and
    (b) work of separation $W$, both times as a function of pulling velocity $v_\textrm{p}$ for a regular Hertzian tip. Adiabatic results from the JKR solution are included for comparison (dashed black lines) as well as fits (dashed brown lines).}
    \label{fig:hertzVE}
\end{figure}

We also considered a Hertzian contact geometry.
The system was modeled numerically with the following parameters:
Radius of curvature $R_\textrm{0} = 1$,
$E^* = 1$, $\tau = 1$, $E_1/E_0 = 100$,
$\Delta \gamma = 2\cdot 10^{-4}$,
$g_\textrm{c} = 1.473\cdot{10}^{-3}$.
With these choices, the Tabor parameter turns out
$\mu_\textrm{T} \approx 2$.
The depinning force is given by the well-known solution by Johnson-Kendall-Roberts (JKR), which was also used to numerically determine the adiabatic work of separation. 
Results are shown in Fig.~\ref{fig:hertzVE}.

As was the case for the cylinder, the work of separation is slightly enhanced for $v \to \infty$ compared to its adiabatic value $W(0)$.
At small velocities, we find an enhancement of both, $F_\textrm{p}$ and $W$, compared to their adiabatic values, which scales roughly with $\sqrt{v_\textrm{p}}$.
For these ``two- plus one-dimensional'' contacts, we did not manage to approach the linear response regime.
Also note that the JKR limits will not be approached exactly for $v \to 0$, since the Tabor parameter of the investigated system was finite.

\bibliographystyle{plain}
\bibliographystyle{unsrt}